\documentclass{aa}

\pdfoutput=1
\usepackage[varg]{txfonts}
\usepackage{amsmath,amssymb}
\usepackage{graphicx} 
\usepackage{times} 
\usepackage{natbib}
\usepackage{lscape}
\usepackage{bigints}
\usepackage[usenames,dvipsnames,table,svgnames]{xcolor}
\usepackage[procnames]{listings}
\usepackage{hyperref} 
\definecolor{linkcolour}{rgb}{0,0.2,0.6} 
\hypersetup{colorlinks,breaklinks,urlcolor=linkcolour,linkcolor=linkcolour,citecolor=linkcolour}
\usepackage{enumerate}

\definecolor{keywords}{RGB}{255,0,90}
\definecolor{listinggray}{gray}{0.9}
\definecolor{lbcolor}{rgb}{0.9,0.9,0.9}
\definecolor{ForestGreen}{rgb}{0.13,0.55,0.13}

\lstdefinestyle{C++} {
  language=C++,
  backgroundcolor=\color{lbcolor},
  basicstyle=\scriptsize\upshape\ttfamily,
  commentstyle=\color{blue},
  classoffset=1,
  morekeywords={cosmobl, DirCosmo, DirLoc, cosmology, Cosmology, par, Object, Galaxy, catalogue, Catalogue, Var},
  keywordstyle=\color{ForestGreen},
  classoffset=2,
  morekeywords={fsigma8, nObjects, setParameters, measure_xi, write_xi},
  keywordstyle=\color{red},
  classoffset=0,
  tabsize=1,
  captionpos=b,
  frame=lines,
  frameround=fttt,
  numbers=left,
  numberstyle=\tiny,
  numbersep=2pt,
  breaklines=true,
  showstringspaces=false
}

\lstdefinestyle{Python} {
  language=Python,
  backgroundcolor=\color{lbcolor},
  basicstyle=\scriptsize\upshape\ttfamily,
  commentstyle=\color{blue},
  classoffset=1,
  morekeywords={Cosmology, pyCosmologyCBL},
  keywordstyle=\color{ForestGreen},
  classoffset=2,
  morekeywords={D_C},
  keywordstyle=\color{red},
  classoffset=0,
  tabsize=2,
  captionpos=b,
  frame=lines,
  numbers=left,
  numberstyle=\tiny,
  numbersep=5pt,
  breaklines=true,
  showstringspaces=false,
  procnamekeys={def,class}
}

\def\apjl{Astrophys. J. Lett.}
\def\apjs{Astrophys. J. Suppl.}

\def\aap{Astron. Astrophys.}

\bibpunct{(}{)}{;}{a}{}{,} 

\DeclareGraphicsExtensions{.eps,.ps,.pdf}

\begin{document}

\title{Cosmological exploitation of cosmic void statistics}

\subtitle{New numerical tools in the CosmoBolognaLib to extract
  cosmological constraints from the void size function}

\author{
  Tommaso Ronconi\inst{\ref{1},\ref{2}}
  \and Federico Marulli\inst{\ref{2},\ref{3},\ref{4}}
}

\offprints{T. Ronconi \\ \email{tronconi@sissa.it}}

\institute{
  SISSA - International School for Advanced Studies, via Bonomea 265, I-34136 Trieste, Italy \label{1}
  \and Dipartimento di Fisica e Astronomia - Universit\`{a} di Bologna, viale Berti Pichat 6/2, I-40127 Bologna, Italy \label{2}
  \and INAF - Osservatorio Astronomico di Bologna, via Ranzani 1, I-40127 Bologna, Italy \label{3}
  \and INFN - Sezione di Bologna, viale Berti Pichat 6/2, I-40127 Bologna, Italy \label{4}
}

\abstract
{ We present new numerical tools to analyse cosmic void catalogues,
  implemented inside the \small{CosmoBolognaLib}, a large set of open
  source C++/Python numerical libraries. }
{ The \small{CosmoBolognaLib} provides a common numerical environment
  for cosmological calculations. This work extends these libraries by
  adding new algorithms for cosmological analyses of cosmic voids,
  covering the existing gap between theory and observations. }
{ We implemented new methods to model the size function of cosmic
  voids, in both observed and simulated samples of dark matter and
  biased tracers. Moreover, we provide new numerical tools to
  construct unambiguous void catalogues. The latter are designed to be
  independent of the void finder, in order to allow a high versatility
  in comparing independent results. }
{ The implemented open source software is available at the GitHub
  repository of the CosmoBolognaLib\thanks{\url{https://github.com/federicomarulli/CosmoBolognaLib}}
  . We also provide a full \texttt{doxygen} documentation and some
  example codes that explain how to use these libraries. }
{}


\authorrunning{T. Ronconi \& F. Marulli}
\titlerunning{New numerical tools for the size function of cosmic voids}

\maketitle


\section {Introduction}

Cosmic voids are large underdense structures that fill a significant
volume fraction of the Universe.  Their great potential as a
cosmological probe for constraining dark energy and testing theories
of gravity has been largely demonstrated at both high and low redshift \citep[e.g.][]{Viel2008, Li2012,
  Clampitt2013, Lam2015, Cai2015, Zivick2015, Barreira2015,
  Massara2015, Pisani2015, Pollina2016.1, Hawken2016, Sahl2015, Sahl2016}. However, there
is no general consensus on how to define these objects. This
represents one of the main issues in their cosmological usage. For
instance, in most cases the size function of cosmic voids detected in
galaxy redshift surveys cannot be directly compared to theoretical
predictions, due to the different void definitions adopted in
observational and theoretical studies \citep{Colberg2005, Sutter2012,
  Pisani2015, Nadathur2015}.

The growing scientific interest in cosmic voids, especially as
cosmological probes of the large-scale structure of the Universe, has
not received an equal coverage from a numerical point of view. Many
cosmic void finders have been developed during recent years
\citep[see][for a cross-comparison of different void
  finders]{Colberg2008}, and some of them are publicly available
\citep[e.g.][]{Neyrinck2008, Sutter2015}. However, there are no
numerical tools available to predict cosmic void statistics and to
perform cosmological analyses.

The {\small CosmoBolognaLib} \citep[hereafter CBL,][]{Marulli2016} is
a large set of open source C++/Python libraries. The main goal of this
software is to provide a common numerical environment for handling
extragalactic source catalogues, performing statistical analyses and
extracting cosmological constraints.  Being a \textit{living project},
new numerical tools are continuously added, both to improve the
existing methods and to provide new ones.

The aim of this work is to upgrade the CBL by including new algorithms
to manage cosmic-void catalogues. We developed a self-consistent set
of functions that allow to compare observed or simulated void
statistics with theoretical predictions.  In particular, we
implement different size function models, that provide the comoving
number density of cosmic voids as a function of their size, redshift,
cosmological parameters, and bias of the sources used to trace the
density field. Moreover, we provide numerical tools to {\em clean} a
generic observed void sample, to be directly compared to theoretical
models.
All the software tools presented in this work are not implemented for a specific void finder, and can be used on the output of any void catalogue.

The CBL provide several algorithms to measure and model the two-point and three-point correlation functions of astronomical sources. The new void algorithms introduced in this work can thus be exploited in a full cosmological pipeline to extract constraints from both the size function and the clustering of cosmic voids.

This paper is organised as follows.  In \S\ref{sec:voidsizefunction}
and \S\ref{sec:voidsizefunctionbias}, we describe the void size
function models implemented in the CBL, for cosmic voids detected both
in dark matter (DM) and in bias tracer distributions.  In
\S\ref{sec:voidcatalogue} we describe the new algorithms to manage
cosmic void catalogues. In \S\ref{sec:conclusions} we draw
our conclusions. One sample Python code that explains how to compute the theoretical void size function is provided in Appendix \ref{sec:vsfexample}, while a C++ sample code to clean a void catalogue is reported in Appendix \ref{sec:c++cata}. We also release a Python script that uses the new CBL algorithms introduced in this work to clean a void catalogue in an automatic way. In Appendix \ref{sec:cpuperformance} we investigate the code performances. Finally, in Appendix \ref{sec:smallerfav} we describe the impact of different assumptions in the cleaning of overlapping voids.  


\section {The size function of cosmic voids in the dark matter distribution}
\label{sec:voidsizefunction}

The distribution of cosmic voids as a function of their size has been
modelled for the first time by \citet{SvdW2004}, with the same
excursion-set approach used for the mass function of DM
haloes \citep{press1974, Bond1991, Lacey1993, Lacey1994,
  Zentner2007}. The key assumption is to define a cosmic void as an
underdensity, originating from the DM density field, that has
grown until reaching the shell crossing.  It can be demonstrated that,
for an initially spherical underdensity, the expanding void shells
cross at a fixed value of density contrast with respect to the
background \citep{Blumenthal1992}.

\begin{figure}
\centering
\includegraphics[width=\columnwidth]{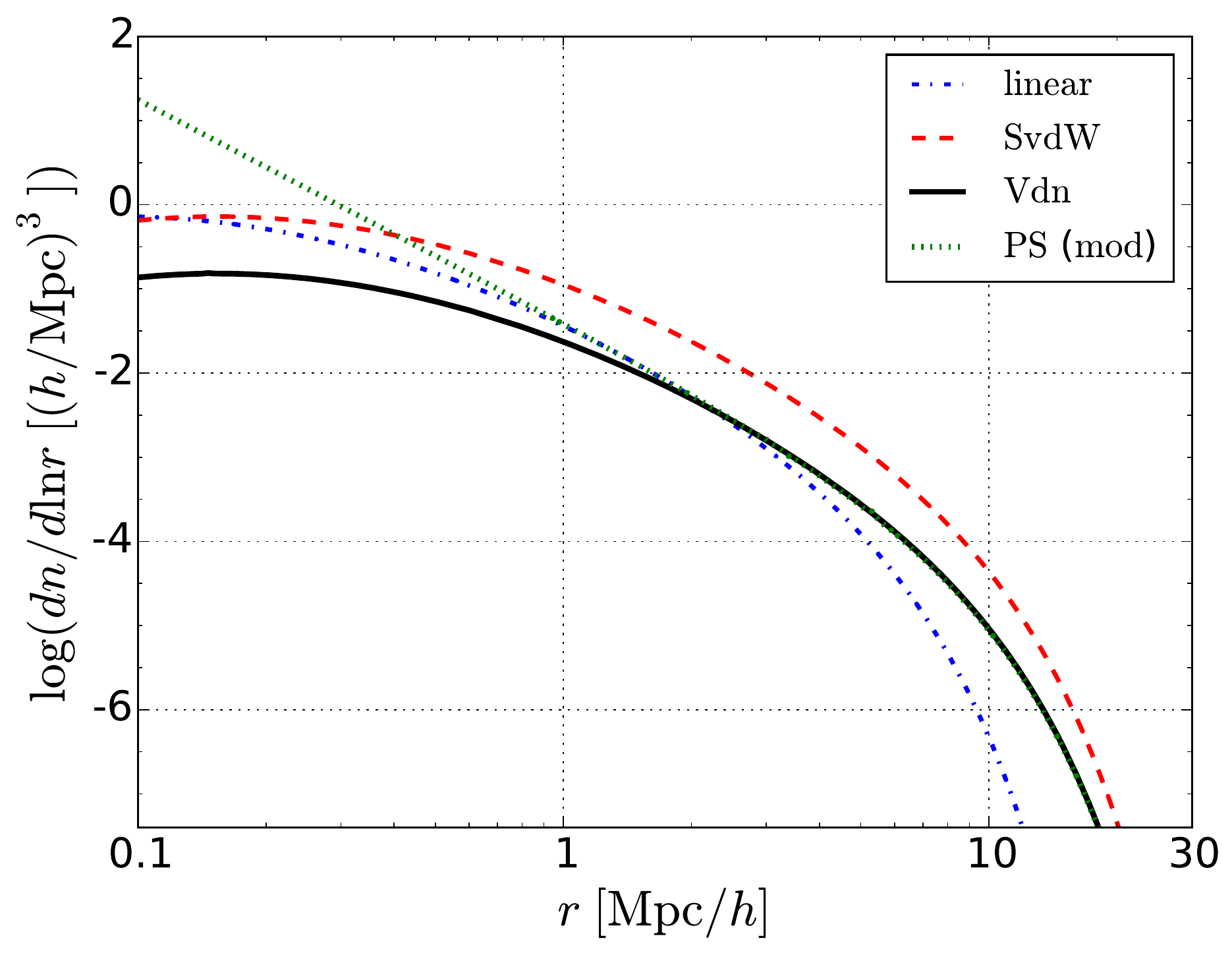}
\caption{The void size function predicted by four different
  models. The blue dot-dashed line represents the linear theory result
  given by Eq. \eqref{eq:vsf03linear}. The red dashed line is the SvdW
  model given by Eq. \eqref{eq:vsf03svdw}. The black solid line is the
  Vdn model given by Eq. \eqref{eq:vsf03vdn}. The green dotted line is
  the void size function obtained from the \citet{press1974} mass
  function modified with Eq. \eqref{eq:vsf05}.}
\label{fig:vsf01}
\end{figure}

We implemented different models to estimate the size function of
cosmic voids in the DM distribution\footnote{The models are
  implemented in the \texttt{cosmobl::cosmology::Cosmology}
  class.}. The implemented software provides full control in the
definition of the cosmological model, that guarantees a wide range of
applicability for cosmological studies.  The excursion-set theory
applied to underdensities \citep[][]{SvdW2004, Jennings2013} predicts
that the fraction of the Universe occupied by cosmic voids is given
by:
\begin{equation}
  \label{eq:vsf01}
  f_{\ln\sigma} = 2 \sum_{j=1}^{\infty}j \pi x^2 \sin(j \pi
  \mathcal{D})\exp\biggl[-\frac{(j \pi x)^2}{2}\biggr]\, ,
\end{equation}
where
\begin{equation*}
  x = \frac{\mathcal{D}}{|\delta_v|}\,\sigma 
\end{equation*}
and
\begin{equation*}
  \mathcal{D} = \frac{|\delta_v|}{\delta_c + |\delta_v|}\, .
\end{equation*}
$\sigma$ is the square root of the variance, computed in terms of the
size of the considered region:
\begin{equation*}
\sigma^2=\frac{1}{2\pi}\int k^2P(k)|W(k,r)|^2\text{d} k \, ,
\end{equation*}
where $P(k)$ is the matter power spectrum, $W(k,r)$ the window
function, and $r$ is the radius of the spherical underdense region
defined as void. Equation \eqref{eq:vsf01} is obtained by applying the
excursion-set formalism with two density thresholds; one positive,
$\delta_c = 1.686$, and one negative, whose typical value is $\delta_v
= -2.71$. The latter is the shell-crossing threshold for
underdensities.  To implement Eq. \eqref{eq:vsf01}, we applied the
approximation proposed by \citet{Jennings2013}, which is accurate at
the $\sim0.2\%$ level:
\begin{equation} 
  \label{eq:vsf02}
  f_{\ln\sigma} (\sigma) =
  \begin{cases}
    \vspace{10pt} \displaystyle
    \sqrt{\frac{2}{\pi}}\,\frac{|\delta_v|}{\sigma}\exp\biggl(-\frac{\delta_v^2}{2\sigma^2}\biggr)
    & \quad x \le 0.276 \\ \displaystyle 2\sum_{j=1}^4 j \pi x^2
    \sin(j \pi \mathcal{D}) \exp\biggl[-\frac{(j \pi x)^2}{2}\biggr] &
    \quad x > 0.276
  \end{cases}
  \ \text{.} 
\end{equation}
Eq. \eqref{eq:vsf02} is used as a kernel of the size function
model. Specifically, we implemented the following three models:
\begin{itemize}
\item the linear model \citep{SvdW2004}:
  \begin{equation}
    \label{eq:vsf03linear}
    \biggl(\frac{\text{d}\,n}{\text{d}\, \ln r}
    \biggr)_{\text{linear}} = \frac{f_{\ln\sigma} (\sigma)}{V(r)}
    \frac{\text{d}\,\ln\sigma^{-1}}{\text{d}\,\ln r} \text{;}
  \end{equation}
\item the \citet{SvdW2004} (SvdW) model:
  \begin{equation}
    \label{eq:vsf03svdw}
    \biggl(\frac{\text{d}\,n}{\text{d}\, \ln r} \biggr)_{\text{SvdW}}
    = \frac{\text{d}\,n}{\text{d}\, \ln r_L}\biggr|_{r_L =
      r/1.72}\ \text{;}
  \end{equation}
\item the volume conserving (Vdn) model \citep{Jennings2013}:
  \begin{equation}
    \label{eq:vsf03vdn}
    \biggl(\frac{\text{d}\,n}{\text{d}\, \ln r} \biggr)_{\text{Vdn}} =
    \frac{\text{d}\,n}{\text{d}\, \ln r_L} \frac{V(r_L)}{V(r)}
    \frac{\text{d}\, \ln r_L}{\text{d}\, \ln r}\ \text{;}
  \end{equation}  
\end{itemize}
where the subscript $L$ indicates values derived in linear theory, and
$V$ is the volume.  Figure \ref{fig:vsf01} compares the void size
functions predicted by these three models.

The void size function can also be predicted directly from the halo
mass function\footnote{The mass function is implemented in the
  \texttt{cosmobl::cosmology::Cosmology} class.} \citep[see][and
  references therein]{Chongchitnan2016}.  Specifically, the
probabilities that the value of the local density contrast smoothed on
scale $R$, $\delta_R$, is above or below a certain threshold
$\delta_v$, are related to each other by the following equation:
\begin{equation}
  \label{eq:vsf04}
  P(\delta_R < \delta_v) = 1 - P(\delta_R > \delta_v)\ \text{.}
\end{equation}
By taking the derivative of Eq. \eqref{eq:vsf04} with respect to $R$,
it results that the halo size function, that is proportional to
$\text{d}\,P(\delta_R>\delta_v)/\text{d}R$, is equivalent to the void
size function, that is proportional to
$\text{d}\,P(\delta_R<\delta_v)/\text{d}R$.  Thus, to obtain the void
size function, we can use the following equation:
\begin{equation}
  \label{eq:vsf05}
  \frac{\text{d} n}{\text{d}\ln r} \equiv \frac{\text{d}\ln
    M}{\text{d}\ln r}\biggl(M \frac{\text{d}n}{\text{d}M}\biggr) = 3 M
  \frac{\text{d}n}{\text{d}M}\ \text{.}
\end{equation}
The last equality comes from the assumed void sphericity.
Equation \eqref{eq:vsf05} gives the comoving number density of voids per
unit effective radius in linear theory, that is analogous to
Eq. \eqref{eq:vsf03linear}, with the difference that it only depends
on one single underdensity threshold, $\delta_v$.  Following the same
line of reasoning used to compute Eqs. \eqref{eq:vsf03svdw} and
\eqref{eq:vsf03vdn}, it is possible to derive the void size function
in the non-linear regime from the halo mass function. As an example,
the green dotted line in Fig. \ref{fig:vsf01} shows a volume
conserving size function obtained from the \citet{press1974} mass
function.  A detailed comparison between the implemented size function
models with measured void number counts extracted from numerical
simulations is provided in \citet{ronconi2017}.


\section {The size function of cosmic voids in the distribution of biased tracers}
\label{sec:voidsizefunctionbias}

To extract unbiased cosmological constraints from the number counts of
cosmic voids detected in galaxy redshift surveys, the tracer bias has
to be correctly taken into account \citep[see e.g.][]{Pollina2016.1}.
Indeed, voids in the DM and in the DM halo density fields do not trace
the same underdensity regions. To account for the effect of tracer
bias, the void size function has to be modified by changing the
shell-crossing underdensity threshold with a scale- and bias-dependent
barrier \citep{Furlanetto2006}.

Even though this solution might work in principle, the analytical
framework is computationally time consuming, and yields underdensity
threshold values that are too low to be applied in realistic tracer
samples. To overcome this issue, we implemented a simpler method to
embed the bias dependency in the underdensity threshold
value. \citet{Pollina2016.2} found that the relation between the
non-linear density contrast of tracers, $\delta_{v,\, \text{tr}}^{NL}$,
and matter, $\delta_{v,\, \text{DM}}^{NL}$, around voids is linear and
determined by a multiplicative constant which corresponds to the value
of the bias parameter of the tracer sample, $b$:
\begin{equation}
  \label{eq:bias01}
  \delta_{v,\, \text{tr}}^{NL} = b\, \delta_{v,\, \text{DM}}^{NL}\ \text{.}
\end{equation}

To recover the linear density contrast of tracers needed to compute the void size function, $\delta_{v,\,\text{tr}}^{L}$, we apply
the fitting formula provided by \citet{Bernardeau1994} to the
non-linear value given by Eq. \eqref{eq:bias01}, as follows:
\begin{equation}
  \label{eq:bias02}
  \delta_{v,\,\text{tr}}^L = \mathcal{C}\, \bigl[1 - (1 + b\,
  \delta_{v,\, \text{DM}}^{NL})^{-1/\mathcal{C}} \bigr]\ \text{,}
\end{equation}
with $\mathcal{C}=1.594$. Both Eq. \ref{eq:bias02} and its inverse have been implemented in the
CBL, to recover the non-linear density contrast that is used to
estimate the void expansion factor, $r(r_L)$.  These built-in
functions can be used to obtain the void size function with the models
described in \S\ref{sec:voidsizefunction}. An application of this
method is presented in \citet{ronconi2017}.


\section {Void-catalogue cleaner}
\label {sec:voidcatalogue}

\subsection{The method}
\label{sec:the method}

One of the main issues in exploiting cosmic voids as cosmological
probes lies in the different void definitions adopted in
observations and theoretical models. In particular, the size function
models described in \S \ref{sec:voidsizefunction} define the cosmic
voids as underdense, spherical, non-overlapped regions that have gone
through shell crossing. To extract cosmological information from void
distributions it is thus required either to use the same definition
when detecting voids from real galaxy samples, or to clean
properly the void catalogues detected with standard methods.

Following the latter approach, we expanded the CBL by implementing a
new algorithm to manage a detected void catalogue to make it directly
comparable to model predictions\footnote{Specifically, we implemented
  a new dedicated catalogue {\em constructor} in the
  \texttt{cosmobl::catalogue::Catalogue} class.}. The algorithm is
used to select and rescale the underdense regions.
The procedure is completely independent of the void finding algorithm used to select the
underdensities, since it only requires the void centre positions.
Optionally, the user can also provide (i) the effective radii, $R_{eff}$, that is, the radii of the spheres having the same volume of the considered structures; (ii) the central density of the regions, $\rho_0$, that is, the density within a small sphere around the centres; and (iii) the
density contrast between the central and the bordering regions,
$\Delta_v$.
Alternatively, the latter two quantities are computed by two
specific CBL functions\footnote{The two functions to compute $\rho_0$
  and $\Delta_v$ are \texttt{cosmobl::compute\_centralDensity()}
  and \texttt{cosmobl::compute\_densityContrast()}, respectively.}, while the former is set to a large enough value (see description below).

To implement our cleaning algorithm, we follow the procedure described
in \citet{Jennings2013}, which can be divided into three steps.
\begin{itemize}
  
\item \textit{First step:} we consider two selection criteria to
  remove {\em non-relevant} objects, that are: \textit{(i)}
  underdensities whose effective radii are outside a given 
  user-selected range $[r_{min},r_{max}]$ ; if the effective radius, $R_{eff}$, is not provided, it is automatically set to $r_{max}$; and \textit{(ii)} 
  underdensities that have a central density higher than $(1 +
  \delta_v^{NL}) \overline{\rho}$, where $\delta_v^{NL}$ is a given
  non-linear underdensity threshold, and $\overline{\rho}$ is the mean
  density of the sample.  We also included a third optional substep to
  deal with ZOBOV-like void catalogues. Specifically, we prune the
  underdensities that do not satisfy a statistical-significance
  criterium, $\Delta_v > \Delta_{v, 0}$, where $\Delta_{v,
    0}$ is a multiple of the typical density contrast due to Poisson
  fluctuations. Threshold values for several $N$-$\sigma$
  reliabilities are reported in \citet{Neyrinck2008}. The user can
  select which of these three sub-steps to run.
  
\item \textit{Second step:} we rescale the underdensities to select
  shell-crossing regions. To satisfy this requirement, we impose the
  same density threshold as the one used by theoretical models
  (Eq. \eqref{eq:vsf01}).  To this end, our algorithm reconstructs the
  density profile of each void candidate in the catalogue, exploiting
  the highly optimised and parallelised chain-mesh algorithm
  implemented in the CBL\footnote{The chain-mesh algorithm is
    implemented in the \texttt{cosmobl::chainmesh} class.}. After this
  step, the void effective radius is computed as the largest radius
  from the void centre which encloses an underdensity equal to
  $\delta_v$.
  
\item \textit{Third step:} finally, we check for overlaps. This last
  step is aimed to avoid double countings of cosmic voids.  The
  collection of regions left from the previous steps are scanned one
  by one, checking for overlappings.  When two voids do overlap, one
  of them is rejected.
  We consider two different criteria, dubbed {\em larger-favoured} and {\em smaller-favoured}.
  In the larger-favoured (smaller-favoured) case, one of the two overlapping voids is rejected according to which of them has the higher (lower) central density ({\em central density selected}) or the lower (higher) density contrast ({\em density contrast selected}).
  The user can
  choose which one of these criteria to apply. Also this step is
  optimised by exploiting the chain-mesh algorithm.
\end{itemize}

This cleaning algorithm returns an object of the \texttt{catalogue} class,
which can be handled with all the already existing CBL methods.

\subsection{Impact of the cleaning steps}
\label{sec:cleaningsteps}

\begin{figure}
\centering
\includegraphics[width=\columnwidth]{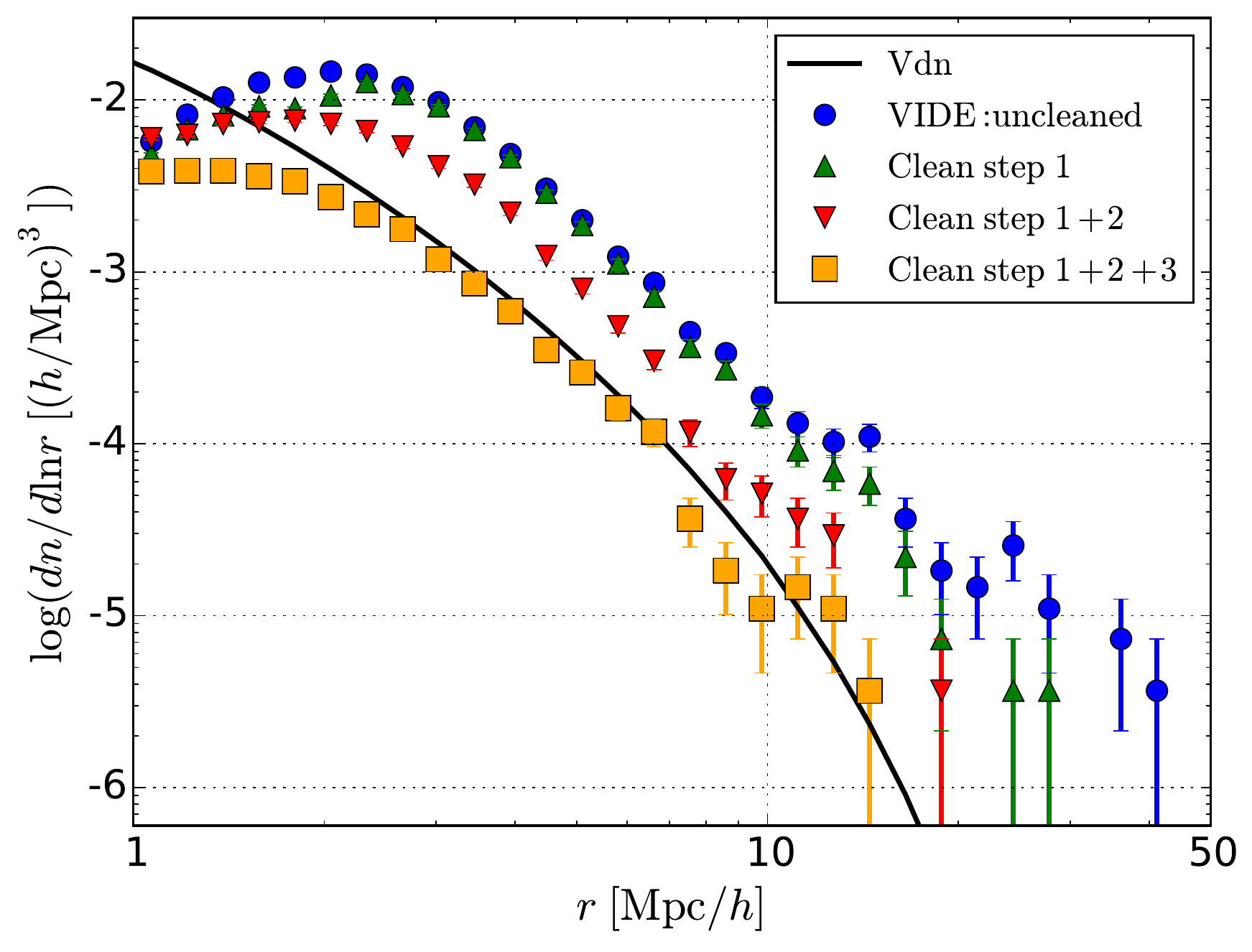}
\caption{ The void size function at 
  the different steps of the cleaning procedure. The blue dots show the distribution of voids detected by
  \small{VIDE} from a
  $\Lambda$CDM N-body simulation. The green triangles are obtained by applying both the $r_{min}$-$r_{max}$
  criterion and
  the central density criterion (first step). 
  The red upside-down triangles
  show the void distribution after having rescaled the voids (second step), while the orange squares are the final output of the cleaning
  algorithm in case the larger-favoured overlapping criterion chosen (third step) is the 
  decreasing central density (for the density contrast criterion see Fig. \ref{fig:alg02} in the Appendices). The black line shows the Vdn model prediction.}
\label{fig:alg01}
\end{figure}

Figure \ref{fig:alg01} shows the void size distribution at the different steps of the cleaning algorithm\footnote{The size distribution has been computed with the
  \texttt{cosmobl::catalogue::var\_distr()} function of the
  \texttt{cosmobl::catalogue::Catalogue} class.}.
The original void sample has been detected with the {\small VIDE} void
finder \citep[][]{Sutter2015} from a $\Lambda$CDM N-body simulation
with $256^3$ DM particles, and side length of $128\ \text{Mpc}/h$. The different symbols show the void size function after each step of the procedure, as indicated by the labels. 

The first selection criterion removes the voids with a radius either smaller than twice the mean particle separation, that for this simulation is $0.5$ Mpc$/h$, or larger than half the boxside length. The effect of this first step is negligible in the size range shown in the Figure. Then the algorithm removes all the voids with central density higher than $0.21\cdot\overline{\rho}$ (second selection criterion of the 
first cleaning step).

The VIDE void radii enclose a compensation wall, that is, an overdense region at the border of the density basin where the matter evacuated
from the inner part accumulates. The second cleaning step of our algorithm provides void radii that are always smaller than the VIDE ones, thus substantially modifying the void distribution, which is shifted to smaller radii. Moreover, the total number count is decreased due to the removal of those underdensities whose radii have been shrinked to zero.

As explained in \S \ref{sec:the method}, the cleaning algorithm provides different options to deal with overlapping voids. With the smaller-favoured criterion (see Appendix \ref{sec:smallerfav}), the size distribution has a loss of power at large radii ($r > \text{Mpc}/h$). On the other hand, voids with radii up to tens of Mpc$/h$ are present when the larger-favoured central density criterion ($\rho_c$) is applied. 

Figure \ref{fig:alg01} shows that all the steps of our cleaning procedure are required to match the theoretical predictions. However, if we clean the overlapping voids according to their density contrast, too many large voids are erased (see Appendix \ref{sec:smallerfav}), leading to an {\em overcleaning}. Better results are found with the 
larger-favoured central density criterion, which provides a good match in the range $2-10\, \text{Mpc}/h$.
The turnover at small radii appears to be in a different location after the cleaning of the void catalogue. In the uncleaned catalogue, the loss of power becomes relevant at around four times the mean inter-particle separation ($\approx 2 \text{Mpc}/h$).
As a result of the rescaling process, small voids in the cleaned catalogue can have radii smaller than the resolution limit. This leads to a plateau in the size function at small radii.

The good agreement between the size distribution of voids in our
cleaned catalogue and the Vdn model demonstrates that there is no need
to fine-tune the model parameters (as has been done in previous works). A systematic comparison of
different size function models is given in \citet{ronconi2017}, who
investigate the possibility of constraining cosmological parameters
from the size distribution of cosmic voids, focusing in particular on
the case of biased samples.


\section{Conclusions}
\label{sec:conclusions}

We implemented a set of new numerical tools to analyse cosmic void
catalogues and model their size distributions. The software is
implemented inside the {\small CosmoBolognaLib}, a large set of open
source C++/Python numerical libraries.
  
Even though some numerical toolkits for the detection of cosmic voids
have been provided in the literature, no cosmological libraries with
void-dedicated tools were available. The CBL provides a set of highly
optimised methods to handle catalogues of extragalactic sources, to
measure statistical quantities, such as two-point and three-point
correlation functions, and to perform Bayesian statistical analyses,
specifically to derive cosmological constraints. This set of libraries
is a \textit{living project}, constantly expanding and upgrading. The
aim of this work was to upgrade the existing software by adding new
numerical tools for a cosmological exploitation of cosmic void
statistics.

We implemented different size function models, for voids detected in
both DM and DM halo density fields. Moreover, we provide a catalogue-cleaning algorithm, that can be used to manage void samples obtained
with any void finder. This point is crucial in order to offer a common
ground to compare results obtained independently. A new void finder
based on dynamical criteria, fully integrated in the CBL, will be
released in the near future \citep{2015MNRAS.448..642E,
  cannarozzo2017}.

The \texttt{doxygen} documentation of all the classes and methods
implemented in this work is provided at the same webpage where the
libraries can be downloaded, together with a set of sample codes that
show how to use this software.


\section*{Acknowledgments}

We thank C. Cannarozzo, L. Moscardini, M. Baldi and A. Veropalumbo for useful
discussions about numerical and scientific issues related to these
libraries.
We also thank M. Baldi for providing the N-body simulations used in this work.
We want to thank the anonymous referee for the useful suggestions that helped to improve this paper.


\bibliography{bib}

\appendix

\section{Size function example}
\label{sec:vsfexample}

The following Python code illustrates how to compute the theoretical
size function of cosmic voids for a given cosmological
model\footnote{We used a similar code to obtain the model size
  distributions shown in Fig. \ref{fig:vsf01}.}.  This is done by
creating an object of the \texttt{Cosmology} class and
then using one of its internal functions to compute the size
function. Alternatively to what is shown in this example, the free model
parameters can be set via a parameter file managed by a dedicated
class (i.e. \texttt{cosmobl::ReadParameters}), as shown in the CBL
webpage.

\begin{lstlisting}[basicstyle=\tiny, style=Python, mathescape, caption={Example of how to compute the SvdW void size function for a given cosmological model.}]
  # import cosmological functions 
  from CosmoBolognaLib import Cosmology 
  
  # define a cosmological model, using default parameters 
  cosm = Cosmology()
  
  # effective void radius
  R = 10.
  
  # redshift 
  z = 0.
  
  # linear underdensity threshold
  del_v = cosm.deltav_L()
  
  # linear overdensity threshold
  del_c = cosm.deltac(0.) 
  
  # size function
  sf = cosm.size_function(R, z, del_v, del_c, "SvdW")
  
  print 'the size function at R =', R, 'Mpc/h and at z =', z,'is', '%.e' % sf, '(h/Mpc)^3'

\end{lstlisting}

\section{Void-catalogue-cleaning example}
\label{sec:c++cata}

The following C++ code shows how to construct a catalogue of spherical
non-overlapped voids which have gone through shell-crossing, and how
to store it in an ASCII file\footnote{We used a similar code to obtain
  the size distributions shown by coloured points in
  Fig. \ref{fig:alg01}.}.

In step I, an ASCII void catalogue\footnote{The provided input
  catalogue has been obtained with {\small VIDE} from a 128 Mpc side
  length $\Lambda$CDM N-body simulation, with $256^3$ DM particles.}
is read by the {\em constructor}, searching for columns containing the
requested attributes.  In step II, we load the original N-body
simulation snapshot into a halo catalogue. Optionally, the user can
exploit a {\em constructor} specifically designed to read binary output
files obtained with GADGET-2 \citep{springel2005gadget2}. Then we
compute the snapshot properties (i.e. volume and mean particle
separation), and use them to store the DM particle positions in a
chain-mesh.  In step III, after computing the central densities of
voids in the input catalogue, we call the catalogue {\em constructor}
which implements the algorithm described in \S\ref{sec:voidcatalogue}.
Finally, we use a built-in function of the CBL to store the new
catalogue in an ASCII file.

\begin{lstlisting}[basicstyle=\tiny, style=C++, mathescape, caption={Example of how to clean a void catalogue to obtain a new catalogue of spherical, non-overlapped voids gone through shell crossing.}]
  // include the header file of libCAT.so
  #include "Catalogue.h"

  // the CosmoBolognaLib and current directories
  string cosmobl::par::DirCosmo = DIRCOSMO
  string cosmobl::par::DirLoc = DIRL;

  int main () {
    
    try {

      // ASCII void catalogue 
      string file_voids_in = cosmobl::par::DirLoc+"../input/vide_void_catalogue.txt";

      // vector containing the variable name list
      vector<cosmobl::catalogue::Var> var_names_voids = {cosmobl::catalogue::Var::_X_, cosmobl::catalogue::Var::_Y_, cosmobl::catalogue::Var::_Z_, cosmobl::catalogue::Var::_Radius_};
      
      // vector containing the columns
      // corresponding to each attribute
      vector<int> columns_voids = {1, 2, 3, 5};

      // catalogue constructor
      cosmobl::catalogue::Catalogue void_catalogue_in {cosmobl::catalogue::_Void_, cosmobl::_comovingCoordinates_, var_names_voids, columns_voids, {file_voids_in}, 1};
      
      // make a shared pointer to void_catalogue_in
      auto input_voidCata = make_shared<cosmobl::catalogue::Catalogue> (cosmobl::catalogue::Catalogue(move(void_catalogue_in)));

      
      // ---------------------------------------
      // ----- build the tracer catalogue ------
      // ---------------------------------------
      
      // binary halo gadget snapshot 
      string file_tracers = cosmobl::par::DirLoc+"../input/tracers_catalogue.txt";

      // vector containing the variable name list
      vector<cosmobl::catalogue::Var> var_names_tracers = {cosmobl::catalogue::Var::_X_, cosmobl::catalogue::Var::_Y_, cosmobl::catalogue::Var::_Z_};
      
      // vector containing the column
      // corresponding to each attribute
      vector<int> columns_tracers = {1, 2, 3};

      // catalogue constructor
      cosmobl::catalogue::Catalogue tracers_catalogue {cosmobl::catalogue::_Halo_, cosmobl::_comovingCoordinates_, var_names_tracers, columns_tracers, {file_tracers}, 1};

      // compute simulation properties
      tracers_catalogue.compute_catalogueProperties();

      // store the mean particle separation 
      double mps = tracers_catalogue.mps();      

      // generate the chain mesh
      // of the inpute tracer catalogue
      cosmobl::chainmesh::ChainMesh3D ChM(2*mps, tracers_catalogue.var(cosmobl::catalogue::Var::_X_), tracers_catalogue.var(cosmobl::catalogue::Var::_Y_), tracers_catalogue.var(cosmobl::catalogue::Var::_Z_), void_catalogue_in.Max(cosmobl::catalogue::Var::_Radius_));

      // make a shared pointer to tracers_catalogue
      auto input_tracersCata = make_shared<cosmobl::catalogue::Catalogue> (cosmobl::catalogue::Catalogue(move(tracers_catalogue)));

      
      // --------------------------------------------
      // ----- build the cleaned void catalogue -----
      // --------------------------------------------

      // compute the central densities
      void_catalogue_in.compute_centralDensity(tracers_catalogue, ChM);

      // the selection criteria,
      // of the First step of the cleaning method, to apply
      // (see Sec. 4)
      vector<bool> clean = {true, true, false};

      // the interval of accepted radii
      vector<double> delta_r = {0.5, 50.};

      // the density threshold
      double threshold = 0.21;

      // the minimum accepted density contrast
      double relevance = 1.57;             

      //catalogue constructor
      cosmobl::catalogue::Catalogue void_catalogue_out {input_voidCata, clean, delta_r, threshold, relevance, true, input_tracersCata, ChM, true, cosmobl::catalogue::Var::_CentralDensity_};

      // store the catalogue in an ASCII file
      var_names_voids.emplace_back(cosmobl::catalogue::Var::_CentralDensity_);
      string cata_out = cosmobl::par::DirLoc+"../output/void_catalogue_cleaned.out";
      void_catalogue_out.write_data(cata_out, var_names_voids);
      
    }

    // catch possible exceptions
    catch (cosmobl::glob::Exception &exc) { std::cerr << exc.what() << std::endl; }
    
    return 0;
  }

\end{lstlisting}

We release also a fully automatic Python pipeline that performs all the cleaning steps of the method. It is available with the latest CosmoBolognaLib release. By setting a self-explaining parameter file, also provided in the package, the user can control each variable of the code. Both the Python script, called cleanVoidCatalogue.py, and the parameter file can be found in the \texttt{Examples} folder of the CosmoBolognaLib package.

\section{Time performances of the cleaning algorithm}
\label{sec:cpuperformance}

We test the time performances of the presented cleaning algorithm as a function of both the density and volume of the input catalogue. The analysis has been performed on one single cluster node with 2 GHz of clock frequency and 64 Gb of RAM memory. 
\begin{figure}
\centering
\includegraphics[width=\columnwidth]{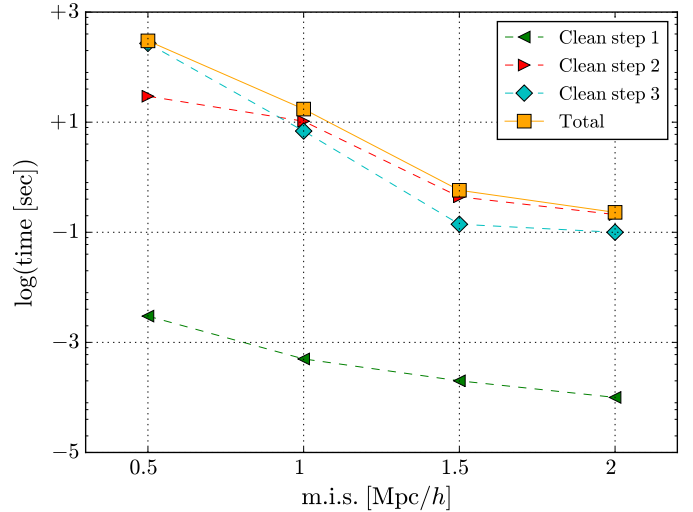}
\caption{ Time performances of the cleaning algorithm as a function of the mean inter-particle separation. The orange dots show the total amount of time spent by the code, while the other symbols refer to each step of the cleaning method, as indicated by the labels.}
\label{fig:time01}
\end{figure} 
Figure \ref{fig:time01} shows the computational time as a function of resolution, that is measured in terms of the mean inter-particle separation (m.i.s.). Both the time spent by each single step of the procedure and the global time are shown. The volume of the simulation used for this analysis is fixed at $V=128^3$ $($Mpc$/h)^3$. 
The first step of the cleaning method is the least time-consuming, while the other two have a similar impact: while the second step contributes the most to the total time at low resolution (high m.i.s.), it becomes less relevant with respect to the third step at higher resolution (small m.i.s.).
The total amount of time spent by the algorithm, as well as that spent by each single step, increases almost exponentially with increasing resolution.

\begin{figure}
\centering
\includegraphics[width=\columnwidth]{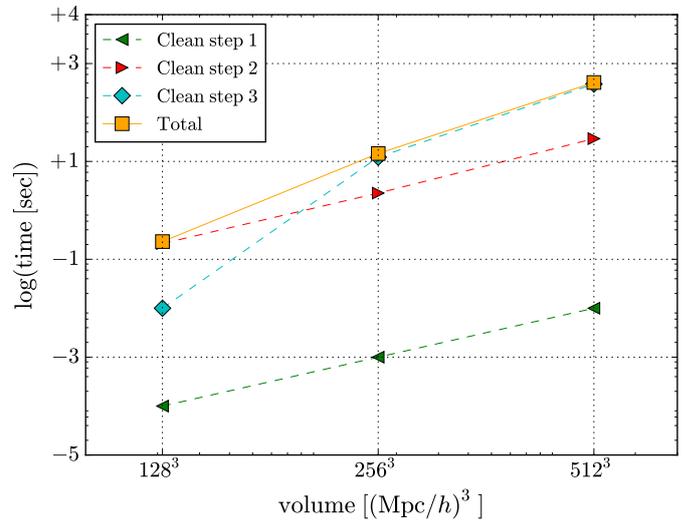}
\caption{The same of Fig. \ref{fig:time01}, but as a function of volume. The resolution is fixed at m.i.s.=$2$ Mpc$/h$.}
\label{fig:time02}
\end{figure}

A similar behaviour is found as a function of volume. The performances of the algorithm at fixed resolution (m.i.s. = $2$ Mpc$/h$) and with varying total volume are shown in Fig. \ref{fig:time02}.
As in the fixed volume case, the first step is the least time-consuming. The third step is less time-consuming at small volumes than the second step, and it takes more time at larger volumes.

The outlined behaviour shows that the time spent by the algorithm increases exponentially with increasing number of voids. In the first case (Fig. \ref{fig:time01}), a larger number of voids are detected at small radii when the resolution is increased, while in the second case (Fig. \ref{fig:time02}) more voids are found at all radii when increasing the volume.

\section{Testing how to clean the overlapping voids}
\label{sec:smallerfav}

\begin{figure}
\centering
\includegraphics[width=\columnwidth]{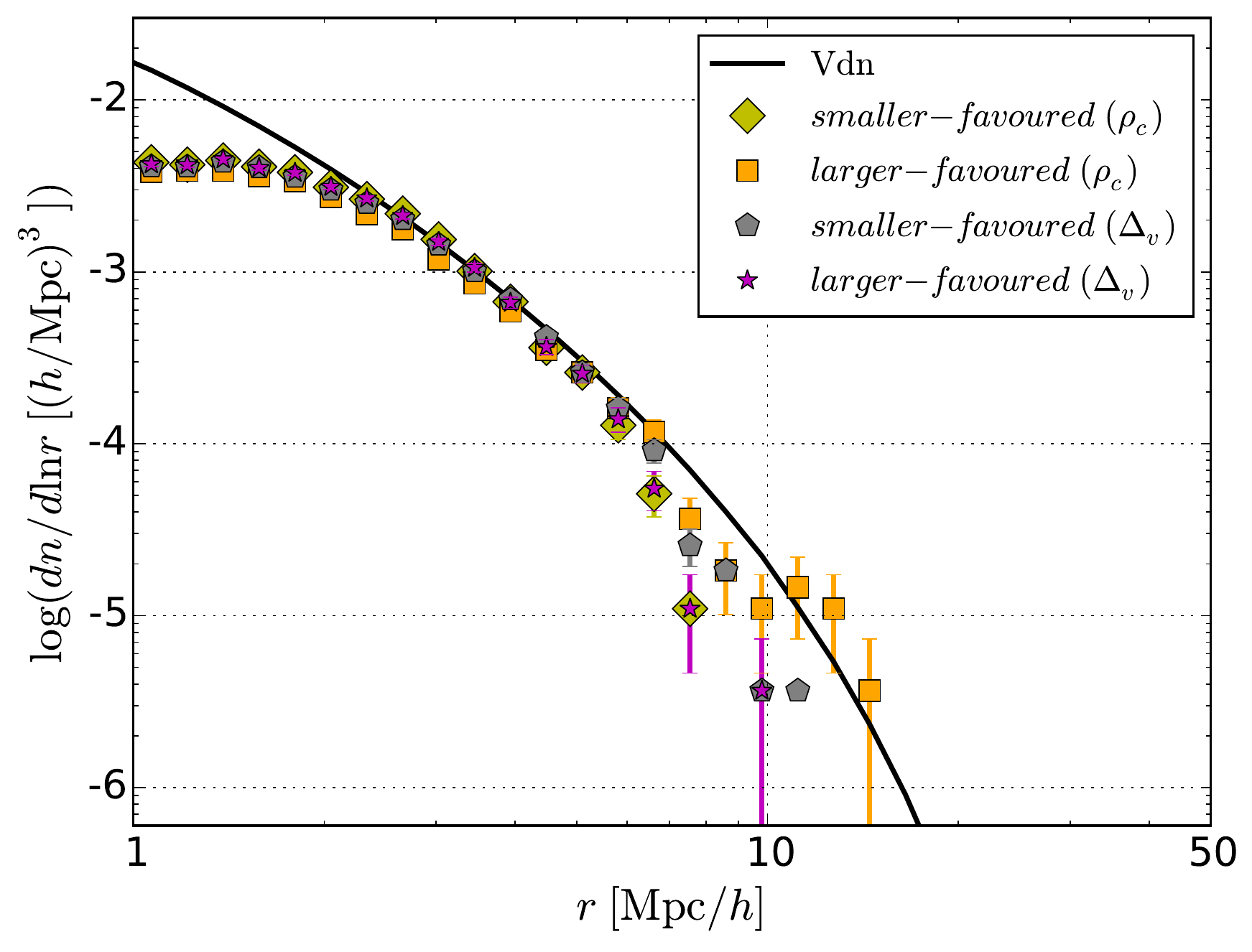}
\caption{ Comparison between the void size functions obtained with the smaller-favoured (central density criteria - yellow diamonds; density contrast criteria - grey pentagons) and the larger-favoured density contrast overlapping criteria (central density criteria - orange squares; density contrast criteria - magenta stars) of the third step of the cleaning method. The black line shows the Vdn model prediction.}
\label{fig:alg02}
\end{figure}

The algorithm presented in this work provides two possible methods to clean the overlapping voids. The choice depends on the purpose of the study the user wants to perform.
Figure \ref{fig:alg02} compares the void size function obtained with all the implemented criteria.

The void size functions obtained with all these different criteria agree reasonably well with the Vdn model.
The two smaller-favoured criteria select deeper density basins, that typically correspond to smaller voids. Such a choice is not convenient for cosmological analyses that aim to extract constraints from the void size function in the largest possible range of radii. Nevertheless, it can be a convenient choice if the goal is to study small voids around the resolution limit of the sample.

\end{document}